\documentclass{iaus}
\usepackage{graphicx}
\title[IAU Symp. 250.~~Massive Stars as Cosmic Engines]{
Massive Stars as Cosmic Engines through the Ages}

\author[Andr\'e Maeder, Georges Meynet, Sylvia Ekstr\"{o}m, Raphael Hirschi, Cyril Georgy]   
{Andr\'e Maeder$^1$,
Georges Meynet$^1$,
Sylvia Ekstr\"{o}m$^1$,\\
Raphael Hirschi$^2$ and
Cyril Georgy$^1$}

\affiliation{$^1$Geneva Observatory, University of Geneva, CH--1290 Sauverny, Switzerland
\\email:andre.maeder@obs.unige.ch;georges.meynet@obs.unige.ch;\\sylvia.ekstrom@obs.unige.ch;
cyril.georgy@obs.unige.ch\\
$^2$Astrophysics, EPSAM, University of Keele\\email:r.hirschi@epsam.keele.ac.uk}

\pubyear{2008}
\volume{xxx}  
\pagerange{}
\jname{Massive Stars as Cosmic Engines}
\editors{}
\begin{document}

\maketitle

\begin{abstract}
Some useful developments in the model physics are briefly presented, followed by model results
on chemical enrichments and WR stars. We  discuss the expected rotation velocities 
of WR stars. We emphasize that the (C+O)/He ratio is a better chemical indicator of evolution for
WC stars than the C/He ratios. With or without rotation,   at a given luminosity   the  (C+O)/He ratios
should be  higher in regions of lower metallicity $Z$. Also,   for a given  (C+O)/He ratio the WC stars in lower $Z$ regions have  higher luminosities.
The   WO stars, which  are likely the progenitors of supernovae SNIc and of some GRBs,  should  preferentially be found in regions of low $Z$ and  be the descendants  of very high initial masses.
Finally, we emphasize the physical reasons why massive rotating low $Z$ stars may also experience 
heavy mass loss.
\end{abstract}

\firstsection 
\section{Introduction}

It may be worth to quote 
a few of the important  findings which have led to the development of our field.  Fifty years ago,
\cite[Peter Conti et al. (1967)]{Conti67}  found that metal deficient stars have higher O/Fe ratios than the solar ratio.   This was the first finding concerning differences of abundance ratios as a function of metallicity, due to a different nucleosynthesis in early stages of galactic evolution.
A year after, \cite[Lindsey Smith (1968)]{Smith68} remarkably found that the various subtypes of WR stars
are differently distributed  in the Galaxy and that some subtypes are missing in the
LMC. The outer galactic regions and the LMC showing the same kind of WC stars (early WC). This was the first evidences of different distributions of massive objects in galaxies. A great discovery
from Copernicus and from IUE is  the mass loss from O--type stars by \cite[Morton (1976]{Morton76}, see also \cite[Donald Morton and Henny Lamers 1976)]{ML76}.
The interpretation of WR stars as post-MS stars resulting from mass loss in OB stars was proposed by 
\cite[Peter Conti (1976)]{Conti76}.

Of course, there are many other big steps which have contributed to our knowledge about massive stars.
It would be misleading to believe that these findings were smoothly accepted as such. For at least a decade,
 there were people claiming that mass loss is an artifact or disputing the status of WR stars, even considering them as pre--MS stars.  These controversies, on the whole, contributed to further checks and investigations which eventually resulted in an increased  strength of the initial discoveries.  

 \section{Improved Model Physics in Massive Star Evolution}

The physics and evolution of massive  stars is dominated by 
mass loss and by  rotational mixing. At the origin of 
both effects, we find the large ratio $T/\rho$ of
temperature to density in massive stars. This   enhances the ratio of radiation 
to gas pressure, which goes like 
\begin{eqnarray}
\frac{P_{\mathrm{rad}}}{P_{\mathrm{gas}}}  \, \sim  \frac{T^3}{\rho} \; .
\end{eqnarray}

\begin{figure}[t]
\begin{center}
 \includegraphics[angle=0,width=7.5cm]{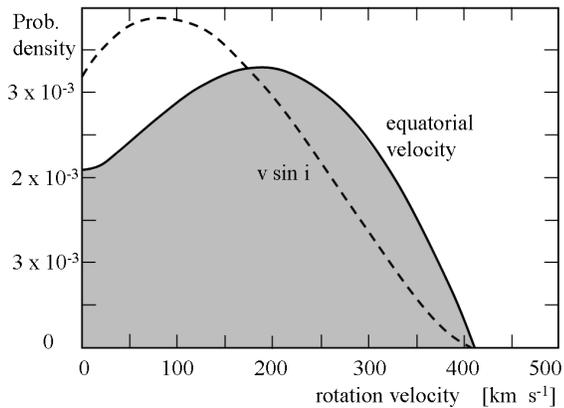} 
 \caption{Probability density by km s$^{-1}$ of rotation velocities for  496  stars with types O9.5 to B8. Adapted from \cite[Huang and Gies (2006a)]{HuangG06a}}
\label{distrv}
\end{center}
\end{figure}

\noindent
The high $T/\rho$  
favors mass loss by  stellar winds.  
A large fraction of OB stars have  high rotational velocities (Fig. \ref{distrv}). 
A high $T/\rho$ also enhances 
rotational mixing either by shear diffusion or meridional circulation, since the coefficient of mixing for a vertical shear $dv/dz$   behaves as
\begin{eqnarray}
D_{\mathrm{shear}}  = \, 2  \,
 \mathcal{R}i_{\mathrm{crit}} \, K \,\frac {\left(dv/dz\right)^2} {N^2_{\mathrm{ad}}} \; . 
 \label{dkg}
\end{eqnarray}

\noindent
$\mathcal{R}i_{\mathrm{crit}}$  is the critical Reynolds number and $N_{ \mathrm{ad}}$ is  the adiabatic Brunt--V\"{a}is\"{a}l\"{a} frequency. The diffusion coefficient 
scales as the thermal diffusivity $K=4acT^3/(3 C_{\mathrm{P}} \,  \kappa \, \rho^2)$. 
Similarly, the velocity of
meridional circulation scales as the ratio $L/M$ of the luminosity to mass.
Thus, the high $T/\varrho$ favors both mixing and mass loss,
 the  account of both effects   brings major
revisions of the model results.

\begin{figure}[t]
\begin{center}
 \includegraphics[angle=0,width=6.7cm]{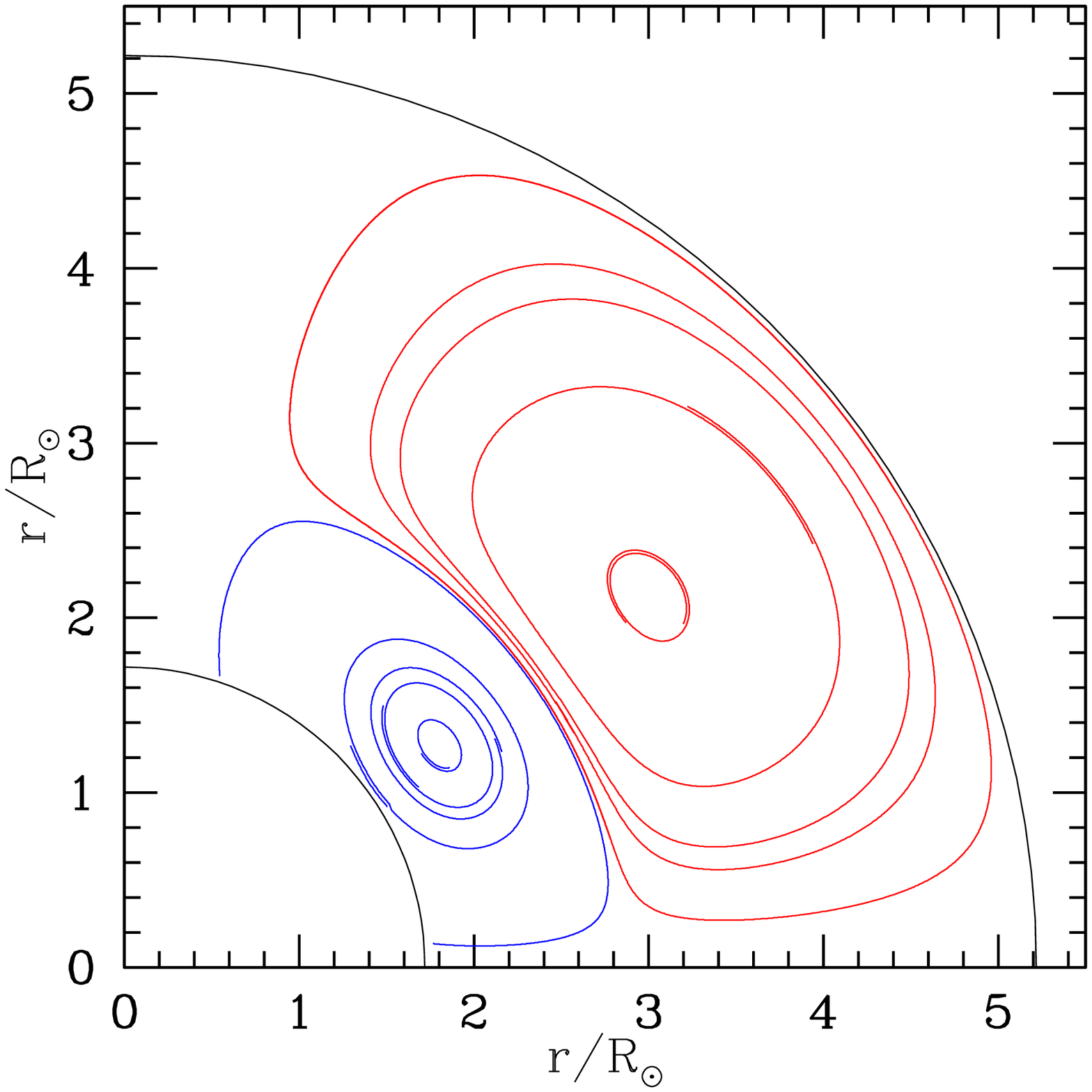} 
 \includegraphics[angle=0,width=6.7cm]{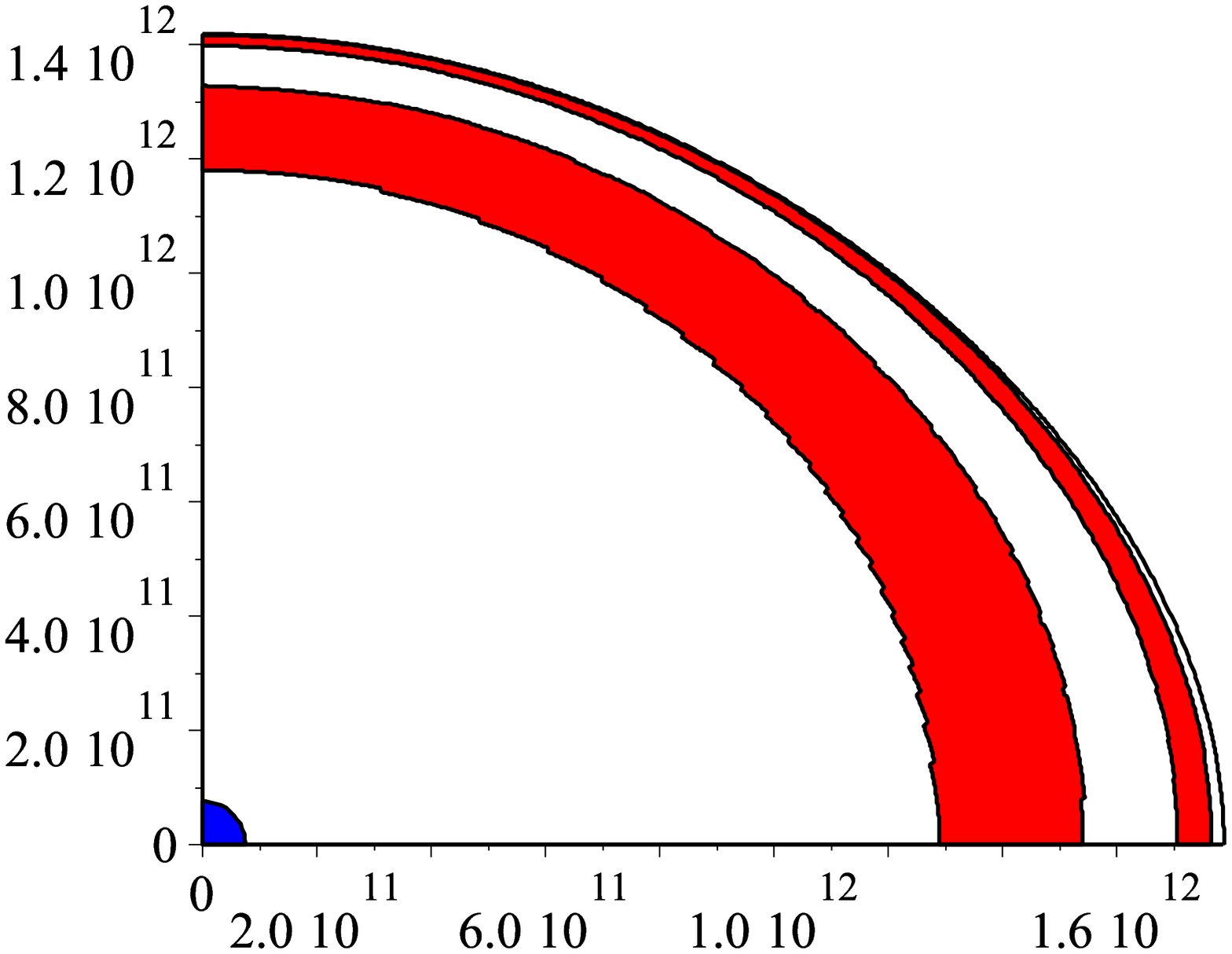} 
 \caption{Left: circulation  currents in a 20 M$_{\odot}$ star in the middle of the H--burning phase.
 The
initial rotation velocity is  300 km s$^{-1}$. The inner loop is raising along the polar axis, while the outer loop,
the Gratton--\"{O}pik circulation cell, is going up in the equatorial plane.
Right: 2--D representation  of the   convective zones (dark areas) as a function of radius   in a model of 20 M$_{\odot}$ with $X=0.70$ and $Z=0.020$ at the end of MS evolution with  ($\Omega/\Omega_{\mathrm{crit}}=0.94$).}
\label{courants}
\end{center}
\end{figure}

 For meridional circulation,  self--consistent solutions were proposed by \cite[Zahn (1992)]{Zahn92}  and
 \cite[Maeder and Zahn (1998)]{MZ98}. 
The transport of chemical elements  obeys a classical diffusion equation. More critical
is the  equation for  the transport of the angular momentum
\begin{eqnarray}
{\partial\over \partial t} (\varrho \, r^2 \sin^2\vartheta \,\Omega)_{r}+{1 \over r^2}{\partial \over \partial r}(\varrho \, r^4\sin^2\vartheta \,U_r\Omega)+{1 \over r\sin\vartheta}
{\partial \over \partial \vartheta}(\varrho \, r^2\sin^3\vartheta \, U_{\vartheta} \Omega) \quad \nonumber \\ [2mm]
= \,{\sin^2\vartheta \over r^2} {\partial \over \partial r}\left(\varrho \, D_{\mathrm{shear}} r^4  {\partial\Omega\over\partial r}\right)+
{1\over \sin\vartheta}
{\partial \over \partial \vartheta}\left(\varrho \, D_{\mathrm{h}}\,  \sin^3\vartheta {\partial \Omega \over \partial \vartheta}\right)\;. \; \; \quad \quad  \; \quad  \quad \;.
\label{eqn5}
\end{eqnarray}
\noindent
It contains both advection terms depending on $U_r$ and $U_{\vartheta}$ the radial and horizontal components of the velocity of meridional circulation
and diffusion terms. $D_{\mathrm{h}}$ is the diffusion coefficient  by the horizontal turbulence. 
The self--consistent solutions allow us to follow the evolution
 of the angular velocity $\Omega(r)$ at each level.   
Many authors 
ignore the advection terms or represent  them by diffusion terms. This is incorrect, since 
the circulation currents may turn in different ways (Fig. \ref{courants} left).

A big question is whether there is a dynamo working in radiative zones, because a magnetic field
would have great consequences on the evolution of rotation by exerting an efficient
 torque able to   reduce  the differential rotation or even to impose uniform rotation.
A dynamo can operate in the radiative zone of a differentially rotating star thanks to a magnetic instability  as shown by \cite[Spruit (2002)]{Spruit02}. Let us consider a star with a shellular rotation
law $\Omega(r)$ with an initially weak poloidal magnetic field  $B_r$, so that the magnetic forces 
are  negligible
(strong initial field leads to solid body rotation).
The radial component is wound up by differential rotation. After a few differential turns,
an azimuthal field  of component $B_{\varphi}$ is present, its strength grows linearly in time 
  and the component  $B_{\varphi}$  dominates over $B_r$. 
 At some  stage, the field $B_{\varphi}$ becomes unstable,  due to Tayler's instability which is the first instability encountered. The instabilities  mainly have  horizontal components, but there is also a small vertical component $l_r$, limited by the action of buoyancy forces. 
This small  radial component of the field is further wound up by differential rotation, which then  
amplifies the toroidal component of the field  up to a stage where dissipation effects would limit its amplitude. In this way, a strong toroidal field develops together with a limited radial field. The horizontal component enforces shellular rotation, while the vertical field component favors solid body rotation. Numerical models by \cite[Maeder and Meynet (2005)]{MagnIII} show
that the field is
  most effective  for transporting angular momentum. 
 The displacement due to the magnetic instability also contribute to enhance the transport of the chemical elements.

In addition to the effects of metallicity on the mass loss rates (see presentations by Vink and  Crowther in this symposium),
the interactions of rotation and stellar winds have many  consequences.
--  Rotation introduces large anisotropies in the stellar winds,
 the polar regions being hotter than the equatorial ones.
 --  The global mass loss rates are increased by rotation.
 -- The anisotropies of the stellar winds allow  a star with strong polar winds to lose
 lots of mass without losing too much angular momentum. At the opposite, equatorial mass loss
 removes a lot of angular momentum.

Let us consider a rotating star with angular velocity $\Omega$
and a non--rotating star of the same mass $M$ at  the same
location in the HR diagram. 
The ratio of their mass loss rates can be  written, see \cite[Maeder (1999)]{MMIV},

\begin{equation}
\frac{\dot{M} (\Omega)} {\dot{M} (0)} \approx
\frac{\left( 1  -\Gamma\right)
^{\frac{1}{\alpha} - 1}}
{\left[ 1 -  \frac{4}{9} (\frac{v}{v_\mathrm{crit, 1}})^2    -\Gamma \right]
^{\frac{1}{\alpha} - 1}}
\; .
\label{momo}
\end{equation}

\noindent

 The ratio $v/v_\mathrm{crit,1}$ is  the ratio of the  rotational velocity $v$
to the critical velocity. If $\Omega = 0 $, $v/v_\mathrm{crit,1}$ is equal to 1.
For a star with a small Eddington factor $\Gamma$, we can neglect $\Gamma$ with respect to unity.
This equation  shows that the effects of rotation on the $\dot{M}$ rates
remain moderate in general. However, for stars close to the Eddington limit, rotation may drastically increase the mass loss rates, in  particularly  for
low values of the force multiplier  $\alpha$, i.e. for  log T$_\mathrm{eff} \leq$
4.30. In  cases where $\Gamma > 0.639$, a moderate
rotation may   make the denominator of (\ref{momo})  to vanish,
indicating large mass loss and instability.

\begin{figure}[t]
\begin{center}
 \includegraphics[angle=0,width=8cm]{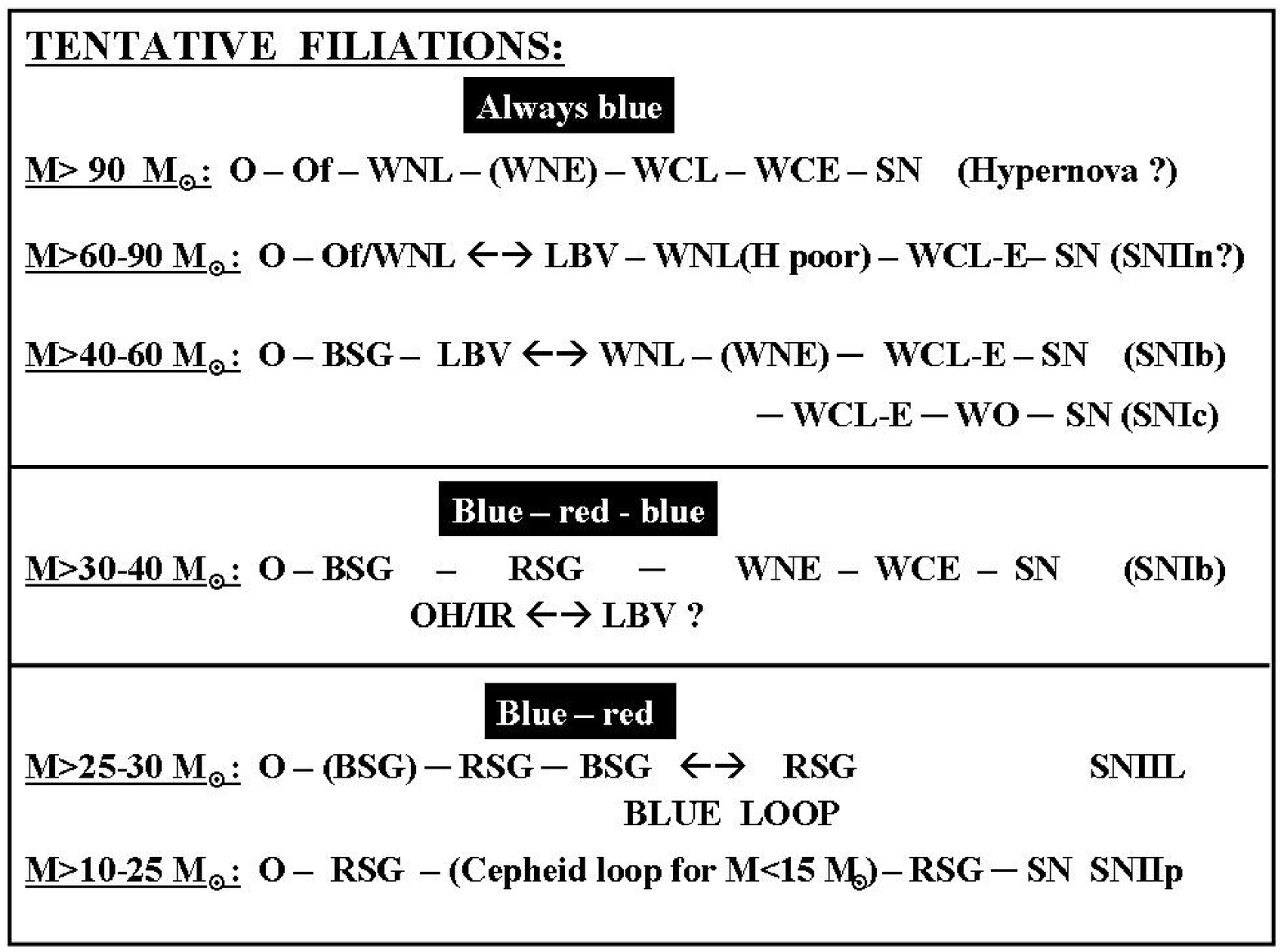} 
 \caption{The filiations between Pop. I massive stars for different mass domains.}
\label{filia}
\end{center}
\end{figure}

Massive O stars have a small external convective envelope
   due to their high luminosity. Rotation amplifies these external convective regions (Fig. \ref{courants} right).
  This occurs  despite the inhibiting effect of the Solberg--Hoiland criterion, because  another more important effect is  present in envelopes: the rotational increase of the radiative gradient $\nabla_{\mathrm{rad}}$ as shown by \cite[Maeder et al. (2008)]{MaederGM08}. These convective zones are likely to  play a large role in the origin of the
clumping of stellar winds.
The matter accelerated in the wind continuously crosses
the convective zones in a dynamical process.
Convection in the outer layers of O--type stars  generates acoustic waves  with periods of several hours to a few days. These waves propagate and are amplified in the winds, which have a lower density.

 
  Most remarkably several of these developments lead to the result that the first stars at very low $Z$ have a very different behavior from the present day massive star evolution.

\section{Evolution with Mass Loss and Rotation}

\subsection{Filiations}

Fig. \ref{filia}  indicates the possible filiations of massive stars of Pop. I, which can be established from  the continuity in the evolution of the chemical abundances, as well as from their properties in star clusters. Globally,  one has  three main cases.

-- For $M>(60-40)$M$_{\odot}$: the high mass loss rates remove enough mass so that stars lose
 their envelopes on the MS or in the blue supergiant stage as LBV. The stars never become
 red supergiants.
 
 -- For about 40 to 30 M$_{\odot}$: the stars only loose a fraction of their envelopes
 on the MS. They further evolve to the red supergiant stage, where mass loss is sufficient to remove their
 envelope, they become bare cores and are observed as  WR stars.

--  Below about 25 to 30 M$_{\odot}$: the stars still experience mass loss,
 however it is not sufficient to  alter the global evolution. The mass loss and rotation may nevertheless still modify the lifetimes  and the chemical compositions.

 The mass limits are uncertain and  depend on  metallicity $Z$.
At different $Z$, some sequences may be absent. The last indicated stage  before supernovae (SN) are usually reached near the end of central He burning. After this stage, the stellar envelopes 
do not  further evolve and  their properties determine the nature of the   SN
progenitors.

\subsection{Chemical Abundances in OB Stars and Supergiants}

\begin{table}[!h]  
 \caption{The largest [N/H] values observed  for different types of stars 
 in the Galaxy, LMC and SMC. The average is equal to about the half of the indicated values.
 See text.} \label{tblabindo}
\begin{center}\scriptsize
\begin{tabular}{lccc}
                &                     &                &               \\
Types of stars     &  [N/H] in Galaxy $\quad$ & $\quad$ [N/H] in LMC $\quad$ & $\quad$ [N/H] in SMC \\
                &                     &                &               \\
\hline 
                &                     &                      &               \\
  O stars                            & 0.8 - 1.0            &   --           &  1.5 - 1.7     \\
  B--dwarfs $M<20$ M$_{\odot}$         & 0.5                 & 0.7 - 0.9      &  1.1            \\
B giants, supg. $M<20$ M$_{\odot}$  & --                  & 1.1 - 1.2      &  1.5            \\
B giants, supg. $M>20$ M$_{\odot}$  & 0.5 - 0.7           & 1.3            &  1.9            \\
                &                     &                      &               \\
 \hline    
\end{tabular}
\end{center}
\end{table}

The chemical abundances offer tests of  stellar physics and evolution. Mass loss, mixing and 
mass exchange in binaries affect surface compositions. 
The removal of the outer layers by stellar winds reveal the inner layers with a composition modified by the nuclear reactions in a beautiful illustration of the effects of the CNO cycles and He--burning reactions. Simultaneously, the internal mixing modifies the surface abundances. Without mixing, there would be no nitrogen enrichment during the MS phase for stars with $M < 60$ M$_{\odot}$. It is only
above this value that mass loss can make the products of the CNO cycle to appear at the stellar surface.

The amplitudes of the enrichments of N/H or N/C  at the end of the MS phase in massive stars is a reference point
telling us the importance of  mixing. The main observations in the Galaxy ($Z\approx 0.02$),
in the LMC ($Z\approx 0.008$) and in the SMC ($Z\approx 0.004$)
 at different $Z$ are summarized in Table \ref{tblabindo}, based on \cite[Herrero (2003)]{Herrero03}, 
\cite[Heap et al. (2006)]{HeapLH06}, \cite[Hunter et al. (2007)]{Hunter07}, 
\cite[Trundle et al. (2007)]{Trundle07}.
We notice  several facts:
\begin{itemize}
\item The N enrichments increase with mass and  evolution.
\item The N enrichments are larger at lower $Z$.
\item  Close to the ZAMS there are both stars
with and without N enrichments. 
\item Away from the ZAMS, but still in the Main Sequence, 
 the  N enrichments are larger  and they are even larger in the supergiant stages.
 In B stars, the He excesses are larger as evolution proceeds on the MS
 and the excesses are greater among the faster rotators as shown by \cite[Huang and Gies (2006a)]{HuangG06a} and \cite[(2006b)]{HuangG06a}.
\end{itemize}

\begin{figure}[t]
\begin{center}
 \includegraphics[angle=0,width=6.7cm]{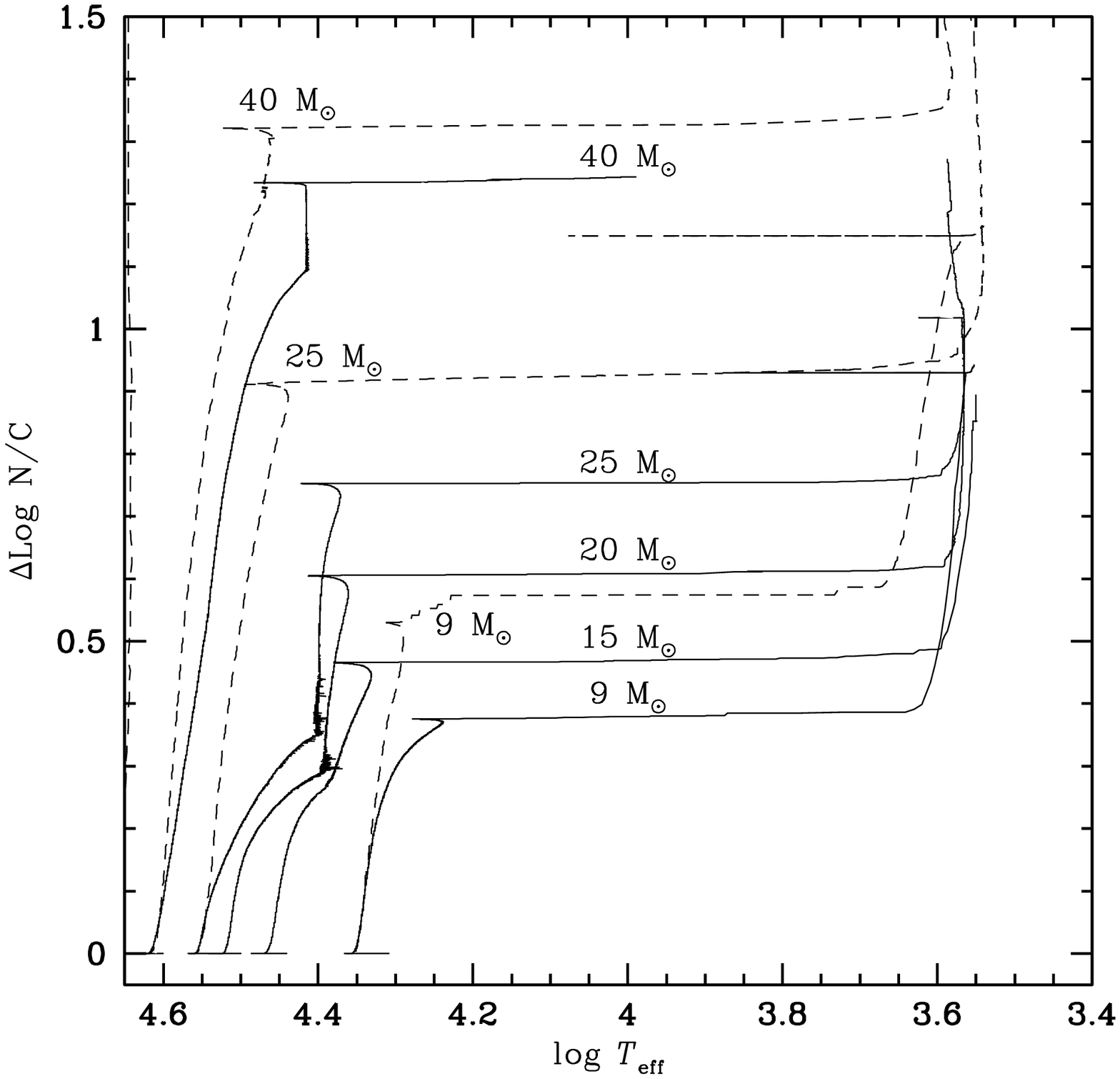} 
 \includegraphics[angle=0,width=6.7cm]{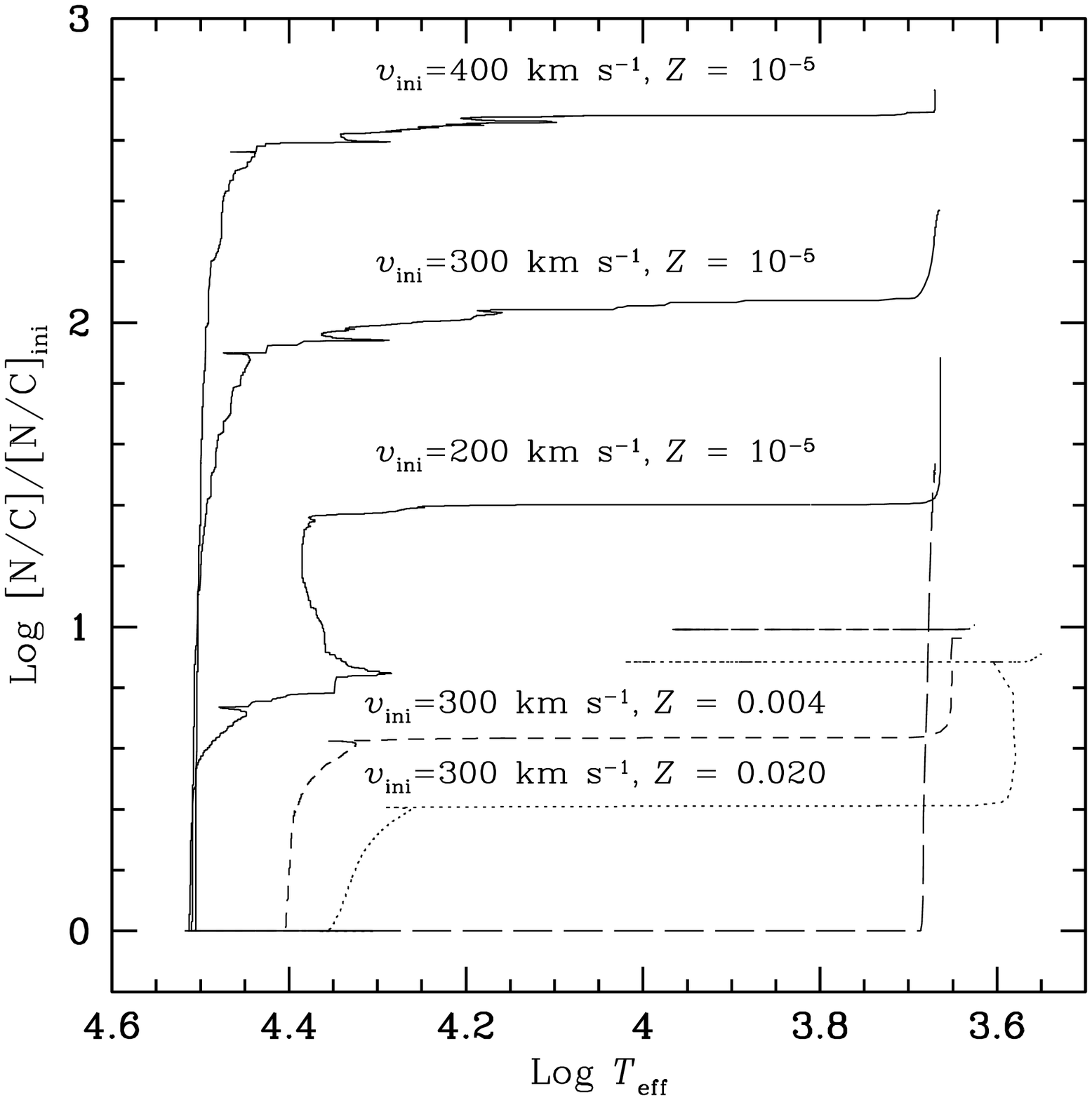}
 \caption{Left: evolution tracks in the plot  $\Delta \log (N/C)$ (change with respect to the initial N/C ratio) vs. $\log T_{\mathrm{eff}}$
         for various initial masses with $Z=0.02$  with initial velocities 300 km s$^{-1}$ (continuous lines). The dashed
         lines show  the tracks with different model assumptions \cite[(Meynet and Maeder 2000)]{MMV}.
          Right: excesses N/C in log scale  for 
a  9 M$_{\odot}$ star at  different  metallicities and  rotation velocities. 
The long--dashed line, at the bottom, corresponds to a non--rotating
9 M$_\odot$ stellar model at $Z=10^{-5}$. From Meynet and Maeder \cite[(2002]{MMVIII}, \cite[2003)]{MMX}.}
\label{nc}
\end{center}
\end{figure}

\noindent
The best  credit should  be given to the sets of data, where the authors carefully distinguish the mass
domains and do not mix in a single plot stars of very different masses. Also, if $\log
 g$ is taken as an indicator of evolution, the rotational effect on the gravity should be accounted for.  Binaries where effects of tidal interactions and mass loss enhancements are possible
 should be separated from single stars. Finally,
 error bars should be indicated. Unfortunately, the non--respect of such wise prescriptions is
 often giving some confusing results.

Fig. \ref{nc} left shows the predicted changes of the $\log(N/C)$ ratios with respect to the initial ratio.  Without rotational mixing (dotted lines), there would be no enrichment until the red supergiant stage. Rotation rapidly increases the N/C ratios
on the Main Sequence, with a level depending on the  velocities. This results from the steep $\Omega$ gradients which produce shear diffusion.
The  predicted N enrichments are in general agreement with  the observed effects. The models  consistently  predict larger N enrichments
with increasing stellar masses and more advanced evolutionary stages.  
If the stars experience blue loops, they have on the blue side of the HR diagram high  N/C ratios   typical of red supergiants. Thus, B supergiants at the same location in the HR diagram  and with the same rotation 
may have very different enrichments.
 The $ v \sin i$   converge toward low
values  during the red phase whatever their initial velocities, thus in the yellow and red phases
stars of almost identical  $ v \sin i$ may exhibit different N/C enrichments.

  Fig. \ref{nc} right
shows the N/C ratios  in models of
rotating stars with 9 M$_{\odot}$ for  $Z$ = 0.02, 0.004 and  10$^{-5}$.
At   $Z$ = 10$^{-5}$ for the 9 M$_{\odot}$ model and  other masses,
there is a large  N/C  increase   by one to two orders of magnitude as shown by Meynet and Maeder \cite[(2002)]{MMVIII}. These very  large enhancements originate from the very steep 
$\Omega$ gradients in rotating stars at low $Z$, which drive a strong turbulent shear diffusion. Consistently with observations in Table \ref{tblabindo}, the lower $Z$ models 
show larger N enhancements. However,
the  observations in the SMC  show  larger N/C ratios than the $Z=0.004$ model of Fig. \ref{nc} right and are in better agreement with relatively lower $Z$ models.

\section{Rotation and Chemistry of WR Stars}

\subsection{Rotation of WR Stars}

\begin{figure}[t]
\begin{center}
 \includegraphics[angle=0,height=8cm,width=10cm]{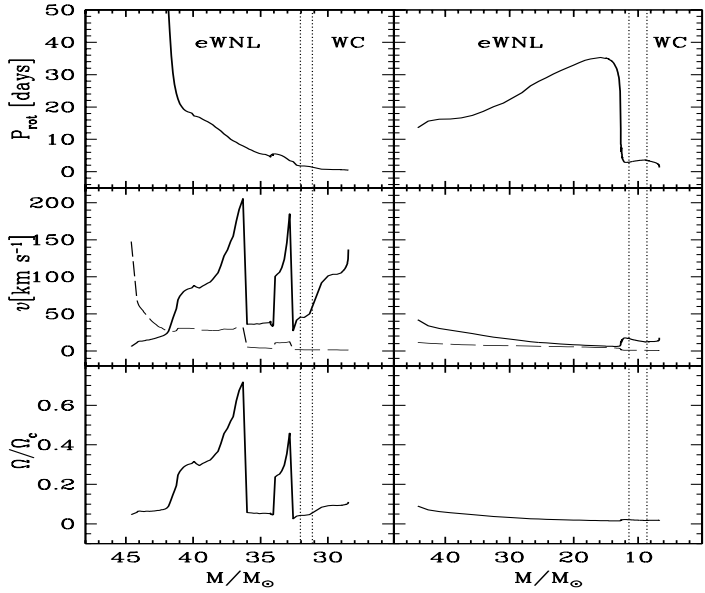} 
 \caption{Evolution as a function of the actual mass 
  of the rotation period, of the surface equatorial velocity
  and of the ratio of the angular velocity to the critical value during
  the WR stage of rotating stars. The long--dashed lines
in the panels for the velocities show the evolution of the radius in solar units.
 Left: the WR phase of a star with an initial
  mass of 60 M$_{\odot}$ with $v_{\mathrm{ini}}= 300$ km~s$^{-1}$ and $Z$ = 0.004. 
 Right: for an initial mass of 60 M$_{\odot}$ with $v_{\mathrm{ini}}= 300$ km~s$^{-1}$
and $Z$ = 0.040. From  \cite[Meynet and Maeder (2005)]{MMXI}.}
\label{V60Z}
\end{center}
\end{figure}

Little is known  on the rotation of WR stars. 
Some information has been recently  obtained  from the  co--rotating regions generating periodic variations in spectral lines (see Chen\'e and St. Louis, this meeting). The velocities are typically lower than about 
50 km s$^{-1}$ in very good agreement with the model predictions.
Fig.~\ref{V60Z} shows 
the evolution during the WR stages of the rotation periods $P =(2\pi/\Omega$), of the
rotation velocities $v$ at the equator and of the fractions 
${\Omega}/{\Omega_{\mathrm{c}}}$ of the angular velocity to 
the critical angular velocity at the surface 
of  star models with an initial mass of
60 M$_{\odot}$ and $v_{\mathrm{ini}}= 300$ km s$^{-1}$ at  $Z = 0.004$ and   0.040.
The evolution  in the WR stages is
fast and the transfer of angular momentum
by meridional circulation is small, thus at this stage the evolution of  rotation 
is dominated by the local conservation of angular momentum unless there is a magnetic field.
The  variations of $v$ and ${\Omega}/{\Omega_{\mathrm{c}}}$ may nevertheless  be fast due to the rapid changes 
of  radius in particular  
when the star loses its last H layers which makes an opacity decrease. The changes of periods are smoother because $v$ and $R$ both decrease at the same pace .

At  solar or higher   $Z$, the expected velocities $v$ are small  with $v< 50 $ km s$^{-1}$. 
The reason is that a large part of the   WNL phase occurs 
during the core H--burning phase, where  
the high mass loss  has time to pump the whole internal angular momentum, so that when the star contracts
to the WC stage there is almost no rotation left.

At lower $Z$, the velocities of WR stars are predicted to be higher, e,g. between 30 and 200 km     
 s$^{-1}$ at $Z=0.004$.
The variations of $v$ and  ${\Omega}/{\Omega_{\mathrm{c}}}$ are also  greater and more rapid when the radius is changing and 
 the break--up limit might be encountered. The reason is that, at 
 low $Z$, the WR stage is not entered during the H--burning phase. Despite mass loss, the inner  rotation
is not killed  due to the  lack of time. 

\subsection{WR Star Chemistry}

\begin{figure}[t]
\begin{center}
 \includegraphics[angle=0,height=9cm,width=9cm]{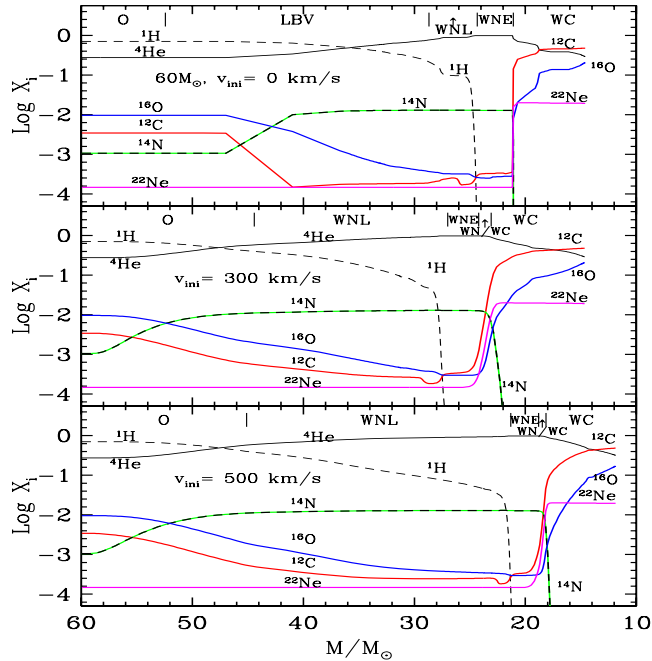} 
 \caption{ Evolution of the surface abundance for  60 M$_{\odot}$ models at $Z=0.02$ for 3 values of the initial  velocities.
         From  Meynet and Maeder\cite[ (2003)]{MMX}.}
\label{dhr60}
\end{center}
\end{figure}

Late WN stars (WNL) generally have H present, with an average value at the surface  $X_{\mathrm{s}} \approx 0.15$, while early WN stars (WNE) have no H left (see \cite[Crowther 2007)]{Crowther07}. In the Galaxy, some WNL stars with weak emission lines have $X_{\mathrm{s}} \approx 0.50$. Other abundance ratios in mass fraction are typically
N/He$=(0.035-1.4)\times 10^{-2}$, C/He$=(0.21-8)\times10^{-4}$ and C/N$=(0.6-6)\times 10^{-2}$.
These values  are very different from the cosmic values   in agreement with model predictions (Fig. \ref{dhr60}). The WN abundances are the values of the CNO cycle at equilibrium, they are   independent of  rotation  and   are a  test  of the nuclear cross--sections. At the transition from WN to WC, the rotating models permit the simultaneous presence
of $^{14}$N, $^{12}$C and $^{22}$Ne enrichments for a short period of time.
This corresponds to the  transition WN/C stars, which show mass fractions of N $\sim1$\% and  C $\sim 5\%$. They represent $\sim$ 4--5 \% of the WR stars. 
Without  rotational mixing, there would be  no WN/C stars, because of the strong chemical discontinuity at the edge of the convective core in the He--burning phase (Fig. \ref{dhr60}). A smooth chemical  transition is needed to produce them in the process of peeling--off as shown by \cite[Langer (1991)]{Langer91bb}.

\subsection{A Fundamental Diagram for WC Stars}

 WC stars have  mass fractions of C between about 10 and 60\%  and  of about 5--10 \% for O, the rest being helium, e.g. \cite[Crowther (2007)]{Crowther07}. 
The variations are smoother in rotating models  (Fig. \ref{dhr60}).  In the WC stage, rotation broadens the range of possible C/He and O/He ratios, permitting the products of He burning to appear at the surface at an earlier  stage of nuclear processing with much  lower C/He and O/He ratios
as suggested by observations. 

 The destruction of $^{14}$N
in the He--burning phase  leads  to the production of $^{22}$Ne,
which  appears at the stellar surface in the WC stage.  The models predict Ne enhancements by a factor 20-30. However, the abundance of the CNO 
elements have been reduced by a factor of $\sim2$  and the Ne abundance has been revised upward by \cite[Asplund et al.  (2004)]{Asplund04}. Thus, a new estimate has to be made. $^{22}$Ne is the daughter
of $^{14}$N, which is itself the daughter of CNO elements. The sum of CNO elements is $X$(CNO)=0.00868,
which essentially becomes $^{14}$N. Since two $\alpha$ particles are added to
$^{14}$N to form $^{22}$Ne, the  abundance of $^{22}$Ne in WC stars should be
\begin{eqnarray}
X(^{22}\mathrm{Ne}) = \frac{22}{14} \,  X(^{14}\mathrm{N}) \, \quad \mathrm{num.} \quad
X(^{22}\mathrm{Ne})= 1.57 \times 0.00868=0.0136 \, . \quad \quad  \quad 
\end{eqnarray}

\noindent
We get a sum of Ne isotopes of about $X(\mathrm{Ne})=0.0156$ compared to  $X(\mathrm{Ne})_{\odot}=0.0020$.
This gives a relative   Ne enhancement by a factor of 8, instead of 20 to 30 with the old abundances.
The factor of 8 is in excellent agreement with the observations by  
\cite[Ignace et al. (2007)]{Ignace07}, who find excesses  of 9.

\begin{figure}[t]
\begin{center}
 \includegraphics[angle=0,width=9.5cm]{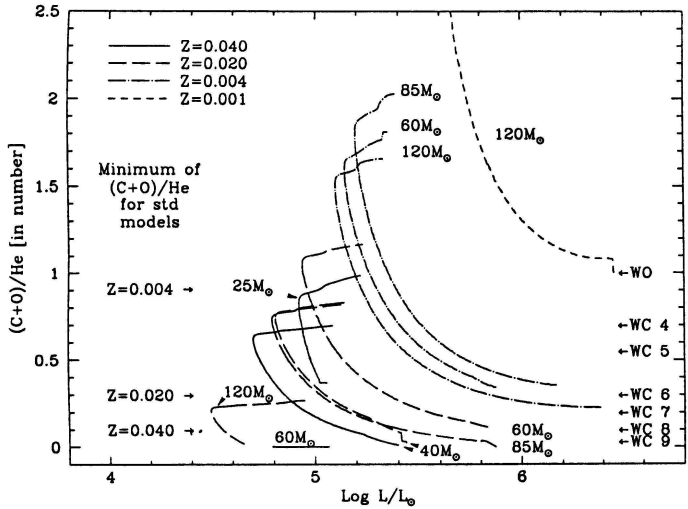} 
 \caption{The (C+O)/He ratios in WC stars as a function of $L$ for different $Z$ and  initial masses, 
         cf. \cite[Maeder and Meynet (1994)]{MaederM94} and
         \cite[Meynet and Maeder (2005)]{MMXI}.}
\label{COHE}
\end{center}
\end{figure}

Fig. \ref{COHE} is a fundamental diagram for the chemical abundances of WC stars. There are versions of this figure  with only mass loss  and also including rotation by \cite[Meynet and Maeder (2005)]{MMXI}. It shows  
the (C+O)/He ratios as a function of the luminosity for WC stars of different initial masses and metallicities.
At  low $Z$, since mass loss is low,  only the extremely massive stars enter the WC stage, thus their luminosities are  high, as well as their 
(C+O)/He ratios. The reason for the high (C+O)/He ratios is that the rare stars which enter the WC 
stage enter it very late in the process of central He burning or even  they do it after He exhaustion.

 At higher $Z$, such as $Z=0.02$ or $Z=0.04$ , due to the higher mass loss rates less massive stars may become WR stars, thus they have lower luminosities (Fig. \ref{COHE}). 
As the mass loss rates are higher, the products of He burning appear at an earlier stage of nuclear processing, i.e. with much lower (C+O)/He ratios.  
These properties seem unavoidable and  they are also present in models with rotation
(rotation  does not alter the  relations  illustrated by Fig. \ref{COHE}).
The coupling  between   $L$, $Z$ and the (C+O)/He ratios   produces the following consequences, see \cite[Smith and Maeder (1991)]{SmithM91}. 

\begin{itemize}
\item  At a given luminosity,   the  (C+O)/He ratios
are higher in regions of lower $Z$.

\item  For a given  (C+O)/He ratio, the WC stars in lower $Z$ regions have much higher luminosities.
\item   WO stars, which correspond to an advanced stage of evolution, should  according to these predictions preferentially be found in regions of low $Z$ and from very high initial masses.
This is a particularly important aspect since WO stars (with little or no He left) are likely the progenitors of supernovae SNIc, a small fraction of which are accompanied by GRBs.
\end{itemize}

These model predictions are  awaiting  observational confirmation. Some of the above trends
have been   discussed  by \cite[Smith and Maeder (1991)]{SmithM91}. However, they rest
on the early results 
by \cite[Smith and Hummer (1988)]{SmithH88}, who suggested  an increase of 
 C/He  from WCL to WCE stars. Further studies have put doubts on this relation
as shown by \cite[Crowther (2007)]{Crowther07}. However, nothing is settled in view of the scarcity of the data on the O abundances.  We emphasize it is essential not to just consider the
C/He ratios, because during the He burning, C is first going up and then going down. Thus, by just 
considering the C data one may get misleading results. The (C+O)/He ratios vary in a monotonic
way and should be preferred as a test of the abundances of WC stars.

\section{Toward the First Stars}

\begin{figure}[b]
\begin{center}
 \includegraphics[angle=0,height=6.0cm,width=8.0cm]{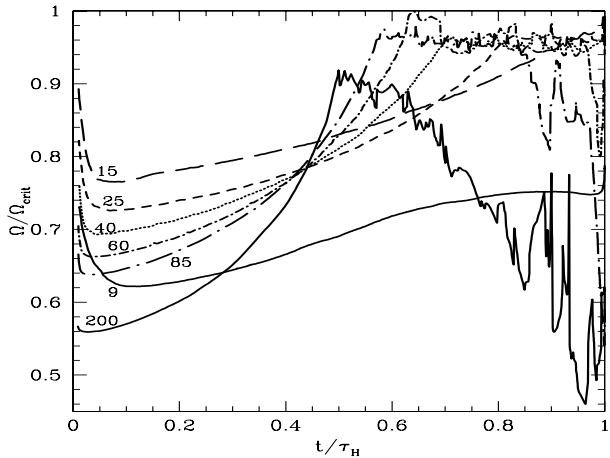} 
 \caption{Examples of the evolution of the ratio of the angular velocity $\Omega$ to the  critical 
 values  as a function of the fraction of the MS lifetime for different initial masses at $Z=0$. From \cite[Ekstr\"{o}m  (2008)]{Ekstrom08}. }
   \label{vrotZ0}
\end{center}
\end{figure}

It is usually considered that mass loss should be small at very low $Z$, such as $Z<10^{-3}$.
This is not necessarily true as  shown by Meynet and Maeder \cite[(2002)]{MMVIII}  and
\cite[Meynet et al. (2006)]{EkstromMM06}. 
There are three different effects intervening.
\begin{itemize}
\item  During the MS phase, the internal coupling of the angular
momentum resulting from Eq. (\ref{eqn5}) is sufficient to transmit some  of  the fast rotation of the contracting 
core to the stellar surface. Thus, for  a large range of initial masses and  $\Omega$ values  the 
low--$Z$ stars reach the critical velocity during their  MS phase  (Fig. \ref{vrotZ0}).  This produces some moderate mass loss during the MS phase.
\item The second effect is due to the self--enrichment of the stellar surface in CNO elements due to
internal mixing.
There is a remarkable interplay \ between rotation, mass loss and chemical enrichments
in low--$Z$ stars.
Low $Z$  implies a weak Gratton--\"{O}pik
circulation, which favors  high  internal $\Omega$--gradients and in turn 
 strong mixing of the chemical elements.  The surface enrichments are very important,  mainly due to the stellar radii being small (the diffusion timescales vary like $R^2$). 
 Then, the high surface enrichments 
in heavy elements, particularly CNO elements,  permits radiative   winds  and mass loss in the He--burning phase of massive  and AGB stars. 
\item The stars with $M< 40$ M$_{\odot}$, depending on rotation, may make blue trips in the HR diagram. If so, the contraction of the convective envelope brings the
surface velocity to critical value and mass loss is enhanced according to Eq. (\ref{momo}).
\end{itemize}

\begin{figure}[t]
\begin{center}
 \includegraphics[angle=-90,width=12cm]{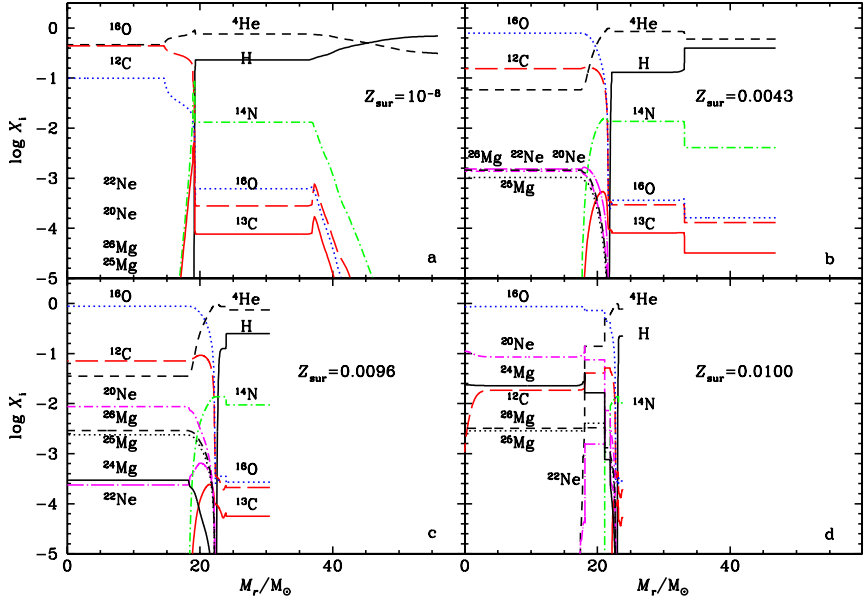} 
 \caption{
Variations of the abundances (in mass fraction) as a function of the Lagrangian mass
within a 60~M$_\odot$ star with
$\upsilon_{\rm ini}$=~800~km~s$^{-1}$ and $Z=10^{-8}$.
 The four panels show the chemical
composition at four different stages at the end of the core He-burning phase:
in panel {\bf a)} the model has a mass fraction of helium at the centre, $Y_{\rm c}$=~0.11
and an actual mass $M$=~54.8~M$_\odot$ - {\bf b)} $Y_{\rm c}$=~0.06,
$M$=~48.3~M$_\odot$ - {\bf c)} $Y_{\rm c}$=~0.04, $M$=~31.5~M$_\odot$
- {\bf d)} End of the core C-burning phase, $M$=~23.8~M$_\odot$. The actual surface metallicity $Z_{\rm surf}$ is indicated in each panel. From \cite[Meynet et al. (2006)]{EkstromMM06}.}
  \label{3070fig6}
\end{center}
\end{figure}

Fig. \ref{3070fig6} illustrates the very strong enrichments in CNO during the He--burning phase
of 60 M$_{\odot}$ stars with an initial $Z=10^{-8}$ and an initial rotation velocity of 800
km s$^{-1}$. From an initial $Z=10^{-8}$, the surface metallicity is brought to
$Z=0.01$. This permits the stellar winds of red supergiants  and AGB stars to remove the
stellar envelopes, especially  as the surface is enriched in C which may produce lots of dust.
The low--$Z$  AGB and massive stars may lose a large fraction of their mass.

These stellar winds  produce very peculiar chemical enrichments of the early galaxies.
The chemical composition of the rotationally enhanced winds of very low $Z$ stars show 
large CNO enhancements  by factors of $10^3$ to $10^7$, together with large excesses of $^{13}$C
and $^{17}$O  and moderate amounts of Na and Al. The excesses of primary N are particularly striking.
When these ejecta from the rotationally
enhanced winds are diluted with the supernova ejecta from the corresponding CO cores, we find [C/Fe], [N/Fe],[O/Fe]
abundance ratios that are very similar to those observed in the C--rich, extremely metal--poor stars 
as shown by \cite[Meynet et al. (2006)]{EkstromMM06}.
Rotating AGB stars and rotating massive stars have about the same effects on the CNO enhancements.
Nevertheless, abundances of s-process elements and the $^{12}$C/$^{13}$C ratios could help us to distinguish between contributions
from AGB and massive stars. As shown by \cite[Chiappini et al. (2006)]{Chiappini06}, these peculiar   enrichments remarkably well account for the initial chemical evolution of the C/O, N/O and  O/Fe    ratios in  the Galaxy.




\begin{thebibliography}{}

\bibitem[Asplund et al. (2004)]{Asplund04}
Asplund, M., Grevesse, N.,  Sauval, A.J. et al.  2004, \textit{A\&A}, {417}, 751

\bibitem[Chiappini et al. (2006)]{Chiappini06}
 Chiappini, C.,  Hirschi, R.,  Meynet, G.,  Ekstr\"{o}m, S.,  Maeder, A.,   Matteucci, F.
 2006, \textit{A\&A}, {449}, 27   

\bibitem[Peter Conti (1976)]{Conti76}
{Conti, P.S.} 1976, \textit{Bull. Soc. Roy. Sci. Liege},
9, 193

\bibitem[Conti et al. (1967)]{Conti67}
{Conti, P.S.,  Greenstein, J.L., Spinrad, H., Wallerstein, G., Vardya. M.S.} 1967, \textit{ApJ},
148, 105

\bibitem[Crowther (2007)]{Crowther07}
 Crowther, P. 2007 \textit{ Ann. Rev. Astron. Astrophys.}, {45}, 177 
 
 
 
 \bibitem[Ekstr\"{o}m  (2008)]{Ekstrom08} 
 Ekstr\"{o}m, S. 2008, \textit{Thesis Univ. Geneva}, in prep. 


\bibitem[Heap et al. (2006)]{HeapLH06} 
{Heap, S.R.,  Lanz, T.,  Hubeny, I.} 2006, \textit{ApJ},  {638}, 409  


\bibitem[Herrero (2003)]{Herrero03} 
{Herrero, A.} 2003,  in \textit{CNO in the Universe}, Eds. C. Charbonnel, D. Schaerer, G. Meynet,
\textit{ASP Conf. Ser.}, {304}, 10 


\bibitem[Huang and Gies (2006a)]{HuangG06a}
 Huang, W.,  Gies, D.R. 2006a, \textit{ApJ}, {648}, 580 
 
 \bibitem[Huang and Gies (2006b)]{HuangG06b}
 Huang, W.,  Gies, D.R. 2006b, \textit{ApJ}, {648}, 591 
 
 


\bibitem[Hunter et al. (2007)]{Hunter07} 
{Hunter, I.,  Dufton, P.L.,   Smartt, S.J. et al.} 2007, \textit{A\&A}, {466}, 277 


\bibitem[Ignace et al. (2007)]{Ignace07}
Ignace, R.,  Cassinelli, J.P.,  Tracy, G. et al.  2007,  \textit{arXiv}, 0707.2770v 


\bibitem[Lamers and Morton (1976)]{ML76}
{Lamers,H.J.G.L.M., Morton, D.C.} 1976, \textit{ApJS}, 32, 715

\bibitem[Langer (1991)]{Langer91bb} 
Langer, N. 1991, \textit{ A\&A}, {248}, 531 

\bibitem[Morton (1976)]{Morton76}
{Morton, D.C.}, 1976, \textit{Bull. American Astr. Soc. }
8, 138


\bibitem[Maeder (1999)]{MMIV}
 Maeder, A. 1999  \textit{A\&A}, {347}, 185
 
  \bibitem[Maeder et al. (2008)]{MaederGM08} 
  Maeder, A.,  Georgy, C., Meynet, G. 2008, \textit{A\&A},  in press 
  
\bibitem[Maeder and Meynet (2005)]{MagnIII} 
{Maeder, A., Meynet, G.} 2005, \textit{A\&A}, 440, 1041 

\bibitem[Maeder and Zahn (1998)]{MZ98} 
{Maeder, A., Zahn, J.P.} 1998,\textit{A\&A}, {334}, 1000 


 \bibitem[Meynet et al. (2006)]{EkstromMM06} 
  Meynet, G., Ekstr\"{o}m, S.,    Maeder, A. 2006, \textit{A\&A}, {447}, 623 
 
\bibitem[(Meynet and Maeder, 2000)]{MMV}
{Meynet, G., Maeder, A.} 2000, \textit{A\&A}, {361}, 101  

\bibitem[2002]{MMVIII} 
{Meynet, G., Maeder, A.} 2002, \textit{A\&A}, {390}, 561 

\bibitem[(2003)]{MMX}
{Meynet, G., Maeder, A.} 2003, \textit{A\&A}, {404}, 975
 
 
\bibitem [Meynet and Maeder (2005)]{MMXI}
{Meynet, G., Maeder, A.} 2005, \textit{A\&A}, {429}, 581
 

\bibitem[Lindsey Smith (1968)]{Smith68}
{Smith, L.F.} 1968, \textit{MNRAS}, 141, 317

\bibitem[Smith and Hummer (1988)]{SmithH88} 
 Smith, L.F.,  Hummer, D.G. 1988,  \textit{MNRAS}, {230}, 511 
 
 \bibitem[Smith and Maeder]{SmithM91}
 Smith, L.F.,  Maeder, A. 1991,  \textit{A\&A} {241}, 77 

\bibitem[Spruit (2002)]{Spruit02}
{Spruit, H.C. }  2002, \textit{A\&A}, {381}, 923  

\bibitem[Trundle et al. (2007)]{Trundle07}
 {Trundle, C.,  Dufton, P.L.,  Hunter, I.  et al.} 2007, \textit{A\&A}, {471}, 625 

\bibitem[Zahn (1992)]{Zahn92} 
{Zahn,J.P.} 1992, \textit{A\&A}, {265}, 115 

\end{thebibliography}
\end{document}